\documentclass{PoS}

\title{Sigma term and strangeness content of the nucleon}

\ShortTitle{Sigma term and strangeness content of the nucleon}

\author{S.\,D\"urr$^{\dagger\,1,2}$, Z.\,Fodor$^{1,2,3}$, J.\,Frison$^{4}$,
T.\,Hemmert$^{5}$, C.\,Hoelbling$^{1}$, S.D.\,Katz$^{1,3}$, S.\,Krieg$^{1,2}$,
T.\,Kurth$^{1}$, L.\,Lellouch$^{4}$, T.\,Lippert$^{1,2}$, A.\,Portelli$^{4}$,
A.\,Ramos$^{\dagger\,4}$, A.\,Sch\"afer$^{5}$, K.K.\,Szab\'o$^{1}$}

\author{\it\phantom{nothing}\\
\normalsize{$^1$Bergische Universit\"at Wuppertal, Gaussstr.\,20,
D-42119 Wuppertal, Germany}\\
\normalsize{$^2$J\"ulich Supercomputing Centre, Forschungszentrum J\"ulich,
D-52425 J\"ulich, Germany}\\
\normalsize{$^3$Institute for Theoretical Physics, E\"otv\"os University,
H-1117 Budapest, Hungary}\\
\normalsize{$^4$Centre de Physique Th\'eorique\footnote{CPT is research unit
UMR 6207 of the CNRS and of the universities Aix-Marseille I, Aix-Marseille II
and Sud Toulon-Var, and is affiliated with the FRUMAM.}, Case 907, CNRS Luminy,
F-13288 Marseille, France}\\
\normalsize{$^5$Universit\"at Regensburg, Universit\"atsstr.\,31,
D-93053 Regensburg, Germany}
}

\author{\phantom{\speaker{S.\,D\"urr and A.\,Ramos}}} 

\abstract{A status report is given for a joint project of the
Budapest-Marseille-Wuppertal collaboration and the Regensburg
group to study the quark mass-dependence of octet baryons in
SU(3) Baryon XPT. This formulation is expected to extend to
larger masses than Heavy-Baryon XPT. Its applicability is
tested with 2+1 flavor data which cover three lattice spacings
and pion masses down to about 190\,MeV, in large volumes. Also
polynomial and rational interpolations in $M_\pi^2$ and $M_K^2$
are used to assess the uncertainty due to the ansatz. Both frameworks
are combined to explore the precision to be expected in a controlled
determination of the nucleon sigma term and strangeness content.}

\FullConference{The XXVIII International Symposium on
Lattice Field Theory - LAT2010\\
June 14-19 2010\\
Villasimius, Sardinia, Italy}

\newcommand{\be}{\beta}

\newcommand{\de}{\delta}

\newcommand{\et}{\eta}

\newcommand{\si}{\sigma}

\newcommand{\ch}{\chi}

\newcommand{\Mpi}{M_\pi}
\newcommand{\Fpi}{F_\pi}
\newcommand{\Mka}{M_K}
\newcommand{\Fka}{F_K}
\newcommand{\Met}{M_\et}
\newcommand{\Fet}{F_\et}
\newcommand{\Mss}{M_{\bar{s}s}}

\newcommand{\pad}{\partial}

\renewcommand{\>}{\rangle}

\newcommand{\mr}{\mathrm}

\newcommand{\Nf}{N_{\!f}}

\newcommand{\fm}{\,\mr{fm}}
\newcommand{\MeV}{\,\mr{MeV}}
\newcommand{\GeV}{\,\mr{GeV}}

\newcommand{\bdm}{\begin{displaymath}}
\newcommand{\edm}{\end{displaymath}}
\newcommand{\bea}{\begin{eqnarray}}
\newcommand{\eea}{\end{eqnarray}}
\newcommand{\beq}{\begin{equation}}
\newcommand{\eeq}{\end{equation}}

\long\def\begincomment#1\endcomment{}

\begin{document}


\section{Introduction}

The sigma term and the strangeness content of the nucleon are
phenomenologically relevant quantities which, however, cannot be directly
measured in experiment.

They are of interest, because they link the pion-nucleon and kaon-nucleon
scattering lengths to the hadron mass spectrum, to the quark mass ratio
$m_s/m_{ud}$, where $m_{ud}\!=\!(m_u\!+\!m_d)/2$ denotes the isospin averaged
first generation quark mass, and -- last but not least -- to the strangeness
content of the nucleon which, in turn, has a say on the importance of quantum
fluctuations.

That the sigma term cannot be directly%
\footnote{Measurable quantities like the $\pi N$ and $KN$ scattering lengths
can be linked to (\ref{def_sigma_udnuc},\,\ref{def_sigma_pinuc}) and
(\ref{def_sigma_kanuc}) by means of formulas from XPT, but this should not
be mistaken as a direct measurement of the sigma term.}
measured in experiment follows from the definition
\beq
\si_{\pi N}\equiv 
\<N(p)|m_u \bar{u}u+m_d \bar{d}d|N(p)\>=
\sum_{q=u,d}m_q\,\frac{\pad M_N}{\pad m_q}
\label{def_sigma_udnuc}
\eeq
\beq
\si_{\pi N}\equiv 
\<N(p)|m_{ud}(\bar{u}u+\bar{d}d)|N(p)\>=
m_{ud}\,\frac{\pad M_N}{\pad m_{ud}}
\label{def_sigma_pinuc}
\eeq
since, in nature, we cannot change the quark masses.
Here we have specified the simplification that emerges in the isospin
limit where the up and down quarks assume a common mass $m_{ud}$.
By contrast, on the lattice we can vary the quark masses, and this opens a
unique opportunity to determine the pion-nucleon sigma term from lattice QCD
datasets with several pion masses.

Provided the lattice simulations include a strange quark (which ours do), one
can also study the kaon-nucleon sigma term and the $\bar{s}s$-nucleon sigma
term
\beq
\si_{KN}\equiv\frac{1}{2}(m_{ud}\!+\!m_s)\<N|\bar{u}u+\bar{s}s|N\>=
\frac{1}{2}(m_{ud}+m_s)
\Big\{
\frac{1}{2}\frac{\pad M_N}{\pad m_{ud}}+ 
\frac{\pad M_N}{\pad m_s} 
\Big\}
\label{def_sigma_kanuc}
\eeq
\beq
\si_{\bar{s}sN}\equiv2m_s\<N|\bar{s}s|N\>=
2m_s\frac{\pad M_N}{\pad m_s}
\label{def_sigma_ssnuc}
\eeq
where in practice it is common to trade (\ref{def_sigma_ssnuc}) for the
strangeness content of the nucleon
\beq
y_N\equiv\frac{2\<N(p)|(\bar{s}s)(0)|N(p)\>}
{\<N(p)|(\bar{u}u+\bar{d}d)(0)|N(p)\>}
\eeq
to which it relates via $y_N m_s/m_{ud}=\si_{\bar{s}sN}/\si_{\pi N}$.
Also the sigma terms are linearly dependent in the isospin symmetric case
(which we will assume in the following, unless stated otherwise), since
$\si_{\pi N}/m_{ud}+\si_{\bar{s}sN}/m_s=4\si_{KN}/(m_{ud}+m_s)$.
The sigma terms have the dimension of a mass (without showing any scheme or
scale dependence), while the strangeness content is a pure number.

On the lattice there are two main strategies to determine the sigma terms.
One option is to stick with the definitions in
(\ref{def_sigma_udnuc}-\ref{def_sigma_ssnuc}), which then leads one to evaluate
$q\bar{q}$ in a nucleon in and out state.
This is technically involved (and noisy), due to quark line disconnected
contributions \cite{Collins:2010gr}.
The second option is to use the Hellmann-Feynman theorem (in a form
adapted to field theory~\cite{Brown:1970dd}) and to determine the sigma
terms from the variation of the nucleon mass as a function of the up/down or
strange quark mass.
In the following, we chose the second option, albeit in the version where one
trades the dependence on $m_{ud}$ and $m_s$ for one in $\Mpi^2$ and
$\Mss^2=2\Mka^2-\Mpi^2$, as this choice avoids renormalization issues.
In other words, we measure $M_N$ for various combinations of $\Mpi^2$ and
$\Mss^2$ (and various lattice spacings $a$ and box sizes $L$), interpolate the
results with ansaetze which we will discuss below, and evaluate the slope of
the interpolation function in the relevant direction at the physical mass
point.
In practice, this interpolation is actually an extrapolation in $\Mpi^2$
(whereas the data have essentially the right $\Mss^2$), and it is clear that
outside the range where we have data the uncertainty on the derivative grows
much faster than the uncertainty on the function itself.

In the following we discuss the details of our ensembles in Sec.\,2, and some
key features of a novel functional ansatz for octet baryon masses derived from
covariant baryon chiral perturbation theory (CBXPT) in Sec.\,3.
First experiences with this formula and, as a complement, with more traditional
polynomial and rational ansaetze are reported in Sec.\,4 and Sec.\,5,
respectively.
The main goal is to give a reliable estimate which precision can be achieved
with our current data, and this together with some preliminary numbers and some
outlook is arranged in Sec.6.

\section{Overview of our ``6\,stout'' dataset and scale setting issues}

The dataset to be used is the one that has been generated for our study of the
hadron spectrum in QCD \cite{Durr:2008zz}, and the overall spirit of the
analysis is the same one as in the $f_K/f_\pi$ paper \cite{Durr:2010hr}.

\begin{figure}
\centering
\includegraphics[width=\textwidth]{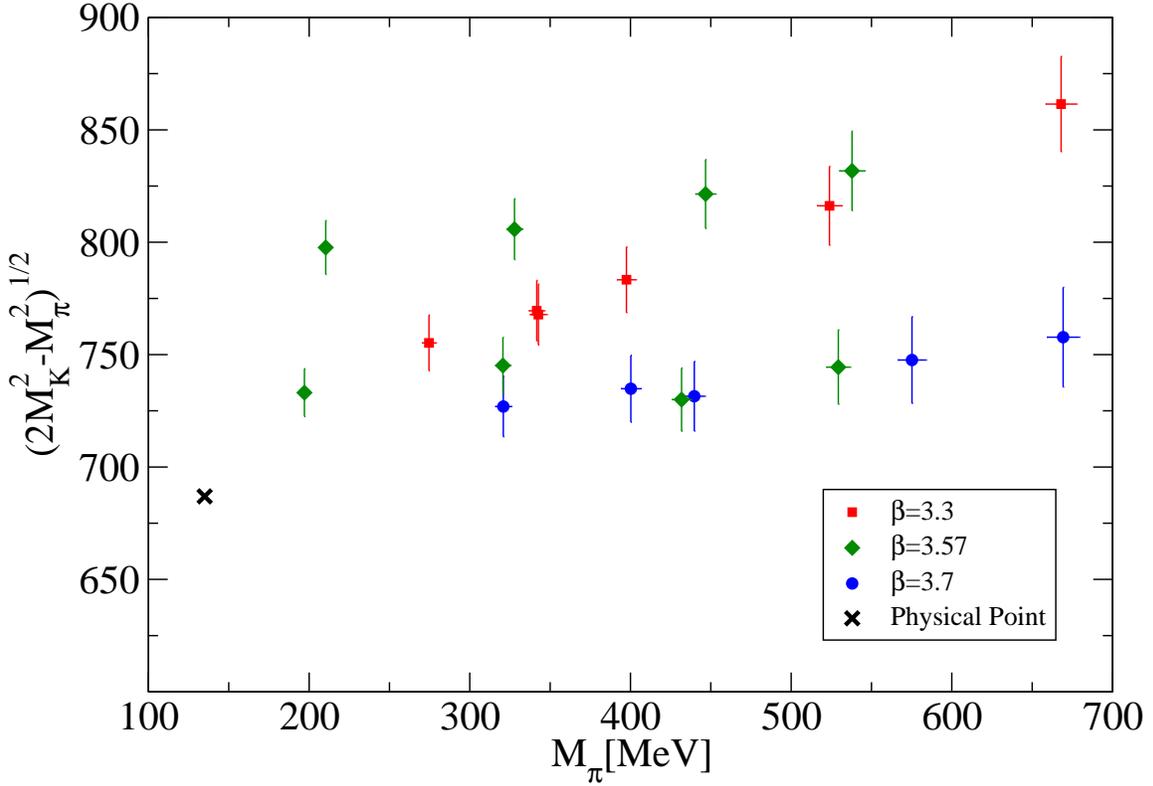}
\caption{Overview of our simulation points in terms of $\Mpi$ and
$\Mss\equiv(2\Mka^2-\Mpi^2)^{1/2}$. The lattice spacings for the three values
of $\be$ are $a\!\simeq\!0.124\fm$ $(\be\!=\!3.3)$, $a\!\simeq\!0.083\fm$
$(\be\!=\!3.57)$, and $a\!\simeq\!0.065\fm$ $(\be\!=\!3.7)$, respectively, and
the physical point is marked with a cross. Error bars are statistical only.}
\label{fig:landscape}
\end{figure}

The ensembles have been generated with two degenerate light quarks and a
separate strange flavor, a scheme commonly referred to as $\Nf\!=\!2\!+\!1$ QCD.
They are based on a Symanzik action with tree-level coefficients for the
gluons and a clover action with 6 levels  of ``stout'' or ``EXP'' smearing
applied to all links which enter the covariant derivative or the clover term.
This combination was found to entail very good scaling properties in standard
hadron observables \cite{Durr:2008rw}.
Note, finally, that our multiply smeared fermion action is as local as the
unsmeared clover quark in the usual sense of locality, i.e.\ $D(x,y)\!=\!0$ for
$||x\!-\!y||\!>\!1$.
Still, in the sense of gauge field locality it has a more extended range:
$\de\!D(x,x)/\de\!U_\mu(y)$ is non-zero for $||x\!-\!y||$ larger than 1 unit,
albeit with a steep fall-off pattern -- see the supplementary online material
of \cite{Durr:2008zz} for details.

Our ensembles include three lattice spacings ($a\simeq0.065,0.083,0.124\fm$)
and the light quark mass $m_{ud}$ is varied such that the pion mass covers the
range from $190\MeV$ to $670\MeV$.
By contrast the strange quark mass $m_s$ is almost the physical one; the
combination $\Mss^2\equiv2\Mka^2-\Mpi^2$ is always in the vicinity of its
physical value.
The physical box size $L$ is such that $\Mpi L$ is in the range of 4 or
larger; this limits finite-volume effects on our hadron masses to an amount
smaller than the statistical fluctuations \cite{Durr:2008zz}.
An overview of these ensembles (to which we refer to as the ``6\,stout'' or
``6\,EXP'' data) in terms of $\Mpi$ and $\Mss\!=\!(2M_{K}^2-M_\pi^2)^{1/2}$
is given in Fig.\,\ref{fig:landscape}.
The smallness of our minimal pion mass $\Mpi^\mr{min}\!\simeq\!190\MeV$ bears
the promise that the extrapolation to the physical point is a relatively mild
one, thus entailing a controllable systematic error in the final result.

All dimensionful quantities in the previous paragraph implicitly build on
knowledge of the lattice spacing $a$.
This brings us to the issue of how the scale is set which, as we shall see,
receives an extra twist when derivatives w.r.t.\ the quark mass are taken,
relative to the case when only spectral quantities at the physical mass point
are calculated.
The physical mass point is the point where any 2 of the 3 ratios that one
may form from $a\Mpi$, $a\Mka$, $aM_X$ (with $X$ being the particle through
which the scale is set, we will consider $X=N,\Xi,\Omega$ below) take on their
physical values.
Apart from the quantity $X$ to be used to set the scale, there is also a choice
regarding the scale setting scheme.
In \cite{Durr:2008zz} we used two such schemes, the ``mass independent scale
setting scheme'' and the ``ratio method''.
In the former case for any given $\be$ the measured $a\Mpi,a\Mka,aM_X$ are
interpolated by a smooth function of the bare quark masses $am_{ud},am_s$
and at the point where $(a\Mpi)/(aM_X),(a\Mka)/(aM_X)$ assume their physical
values the interpolated $aM_X$ is identified with $a$ times the physical value
of $M_X$; this yields the lattice spacing $a$ for \emph{all} ensembles with a
common coupling parameter $\be$.
In the latter case dimensionless ratios are formed on a per ensemble basis,
and these ratios are interpolated with a smooth function of $am_{ud},am_s$,
and read off at the point where $(a\Mpi)/(aM_X),(a\Mka)/(aM_X)$ assume their
physical values.
Effectively this means that the scale is set for each ensemble individually;
the lattice spacing $a$ depends on the combination $(\be,am_{ud},am_s)$.
While these two scale setting schemes yield identical results (at the physical
mass point) for all spectral quantities \cite{Durr:2008zz}, there is a slight
subtlety if sigma terms are evaluated via the Hellmann-Feynman theorem.

For definiteness, let us consider the pion nucleon sigma term at a fixed
physical value of $m_s$, as a function of $m_{ud}$.
The way it is calculated on the lattice amounts to the factorization
\beq
\si_{\pi N}(m_{ud})=m_{ud}\,\frac{\pad M_N}{\pad m_{ud}}\simeq
\Mpi^2\,\frac{\pad M_N}{\pad\Mpi^2}=
\Big[M_X\Big]_\mr{PDG}\,
\Big[\frac{\Mpi^2}{M_X}\,\frac{\pad M_N}{\pad\Mpi^2}\Big]_\mr{latt/phys\!-\!pt}
\label{factorization}
\eeq
where the apparent dependence on the scale setting channel $X$ actually boils
down to cut-off effects at the physical mass point (which then disappear in
the continuum limit), while it seems there is an ambiguity in the last bracket
for which the continuum limit does not provide a remedy.
The point is that the derivative in the last bracket makes reference of the
true scale \emph{away} from the physical mass point, and this is, as we have
just seen, ambiguous.
Perhaps it is easiest to discuss an extreme case first.
If we set the scale through the nucleon, then the pion nucleon sigma term with
the ``ratio method'' is --~by definition~-- zero, i.e.\
$\si_{\pi N}^{N,\mr{rat}}\!=\!0$.
And if we set the scale through the omega, the pion nucleon sigma term with
the ``ratio method'' evaluates to the same amount as the difference between
the pion nucleon sigma term and the pion omega sigma term in the ``mass
independent scale setting scheme'' would, i.e.\ $\si_{\pi N}^{\Omega,\mr{rat}}=
\si_{\pi N}^\mr{misss}-\si_{\pi\Omega}^\mr{misss}$.


Taking a look at (\ref{def_sigma_udnuc},\,\ref{def_sigma_pinuc}) reveals the
origin of this apparent discrepancy.
The matrix elements refer exclusively to the physical mass point, and can be
evaluated without any ambiguity (at least with a Ginsparg-Wilson type action).
The not-so-innocent part is the derivative w.r.t.\ the quark mass taken in the
last equalities.
Here it is (implicitly) assumed that the lattice spacing does not change as
the quark mass is varied, and this is why we must assume --~for consistency
reasons~-- the ``mass independent scale setting scheme'' when evaluating sigma
terms, on the lattice, via the Hellmann-Feynman theorem.
In other words, the requirement to attribute a common lattice spacing $a$ to
all ensembles with one coupling $\be$ comes about through the specific
transcription of the observable of interest; in general there is no restriction
on the scale setting scheme.
Still, it holds true that the QCD $\be$-function depends, for asymptotic
coupling (i.e.\ for small enough lattice spacing), only on the \emph{number} of
active flavors, not on their masses.
Accordingly, the ``mass independent scale setting scheme'' is always applicable,
since it stipulates a property at finite lattice spacing which, due to
asymptotic freedom in QCD, must hold at arbitrarily small lattice spacing.

\section{Octet masses in CBXPT}

Given the discussion in the previous sections, it is clear that a controlled
determination of the pion nucleon sigma term $\si_{\pi N}$ relies on accurate
measurements of the nucleon mass at various pion and kaon mass points, and a
smooth ansatz $M_N=M_N(\Mpi^2,\Mss^2)$ which is valid in the entire regime from
the heaviest datapoint included in the fit down to the physical pion mass (and
eventually down to the chiral limit, if one wishes to determine chiral
low-energy constants).

In the present contribution we focus on $\si_{\pi N}$ at the physical mass
point.
For this purpose it is fully sufficient to come up with an analytic (i.e.\
polynomial or rational) expression for $M_N(\Mpi^2,\Mss^2)$ and an analogous
expression for $M_X(\Mpi^2,\Mss^2)$ for the state $X$ that is used to set the
scale, since --~in the interval between $\Mpi^\mr{phys}$ and $\Mpi^\mr{max}$ of
the dataset~-- QCD is an analytic function of the quark masses.
Preliminary results from such an approach will be reported in Sec.\,5 below.

In a further perspective it is clear that we will not be able to resist the
temptation of testing the host of predictions by chiral perturbation theory
(XPT) of how different quantities relate to each other.
In XPT the quantities connect through their behavior in the (2-flavor or
3-flavor) chiral limit, and this is the reason why, with this goal in mind, the
ansatz must be good all the way down to zero quark mass.
The $\Nf$-flavor chiral limit of QCD is dominated by the logarithmic
singularity induced by the spontaneous breaking of the flavor $SU(\Nf)_A$
symmetry and the dynamics of the pertinent pseudo-Goldstone bosons.
For observables built from pseudoscalar mesons the consequences have been cast
into a valid form in the seminal papers by Gasser and Leutwyler
\cite{Gasser:1983yg,Gasser:1984gg}.

For observables that include baryons the task is significantly more involved,
and we attempt to give a short explanation why there is a plethora of different
versions of chiral perturbation theory for baryons.
Ultimately, they all rest on Weinberg's power counting theorem
\cite{Weinberg:1978kz}, but the basic difficulty is that the quadratic
divergence in self-energies, brought about by a meson loop attached to a given
line, cannot be traded for a logarithmic divergence, as is the case in the
dimensionally regulated setup of meson XPT.
The original approach taken by Gasser, Sainio and Svarc \cite{Gasser:1987rb}
(``OBXPT'') is to live with this fact and to pay a certain price in terms of a
slowly converging series away from the chiral limit.
The heavy baryon approach by Jenkins and Manohar \cite{Jenkins:1990jv}
(``HBXPT'') treats the nucleon as a non-relativistic particle, and in this
approach it was shown that the $\Delta$ can be introduced as an explicit
degree of freedom and that this improves convergence
\cite{Bernard:1992qa,Hemmert:1997ye}.
The covariant approach taken by Becher and Leutwyler \cite{Becher:1999he}
(``CBXPT'') aims at establishing relativistic covariance again, via infrared
regularization.
To the best of our knowledge, in this framework other octet members have not
been included as explicit degrees of freedom yet.

In short one can say that the HBXPT approach is very successful for those
observables where the physics is completely dominated by light (i.e.\ physical
or lighter) pions which are treated as the pseudo-Goldstone mesons in QCD with
2 light flavors.
Extensive work in the nineties showed that applications of HBXPT to baryon
observables that depend on the strange quark flavor lead, in general, to large
(i.e.\ unnatural) cancellations between leading order and next-to-leading order
contributions (see e.g.\ \cite{Borasoy:1996bx,Donoghue:1998bs,Borasoy:1998uu}).
During the first decade of the new millennium, this insight was painfully
re-discovered by the lattice community, as several collaborations attempted
chiral extrapolations ($m_{ud}\to0$ at fixed $m_s$) of $2\!+\!1$ flavor baryon
data, using only leading-order HBXPT formulas.
This triggered a search for ``better'' (i.e.\ more practical) recipes, for
instance
\begin{itemize}
\itemsep-2pt
\item
finite-range regulator extensions of HBXPT to include or model higher order
effects (see e.g.\ Amherst group \cite{Donoghue:1998bs} and Adelaide group
\cite{Leinweber:1999ig,Young:2002ib})
\item
taming higher order effects by adding a host of decuplet contributions in
HBXPT/NRSSE (see e.g.\ LHP Collaboration \cite{WalkerLoud:2008bp} and others
\cite{Tiburzi:2008bk})
\item
resort to polynomial/rational fit functions in $\Mpi^2\propto m_{ud}$ (e.g.\
us \cite{Durr:2008zz,Durr:2010hr} and other groups)
\item
resort to polynomial/rational fit functions in $\Mpi\propto m_{ud}^{1/2}$
(see \cite{WalkerLoud:2008bp,WalkerLoud:2008pj}).
\item
modify integration contour in CBXPT approach (cf.\ Dorati-Gail-Hemmert
\cite{Dorati:2007bk})
\end{itemize}
but a fair review of these developments is clearly beyond the scope of this
contribution.

\begin{table}[tb]
\centering
\begin{tabular}{|c|cccc|}
\hline
$g_{M\!B}$ & $N$ & $\Lambda$ & $\Sigma$ & $\Xi$ \\
\hline
$\pi$ & $\frac{3}{4}(D\!+\!F)^2$ & $D^2$ &
        $\frac{1}{3}(D^2\!+\!6F^2)$ & $\frac{3}{4}(D\!-\!F)^2$\\
$K$   & $\frac{1}{6}(5D^2\!-\!6DF\!+\!9F^2)$ & $\frac{1}{3}(D^2\!+\!9F^2)$ &
        $(D^2\!+\!F^2)$ & $\frac{1}{6}(5D^2\!+\!6DF\!+\!9F^2)$ \\
$\et$ & $\frac{1}{12}(D\!-\!3F)^2$ & $\frac{1}{3}D^2$ &
        $\frac{1}{3}D^2$ & $\frac{1}{12}(D\!+\!3F)^2$\\
\hline
\end{tabular}
\caption{Summary of the baryon-meson-baryon couplings $g_{MB}$ in terms of the
low-energy constants $D,F$.}
\label{tab:gmb}
\end{table}

In the following, we are going to explore the suitability of one such formula,
which belongs to the last point in this list.
It has been worked out by one of us (TH) and involves the function
\beq
H(X^2)\equiv-\frac{X^3}{4\pi^2}\,
\bigg\{\sqrt{1\!-\!\frac{X^2}{4M_0^2}}\,\arccos\Big(\frac{X}{2M_0}\Big)+
\frac{X}{4M_0}\log\Big(\frac{X^2}{M_0^2}\Big)\bigg\}
\label{hemmert0}
\eeq
where $X$ will be a meson mass.
Furthermore, it builds on $SU(3)$ relations among the various
baryon-meson-baryon couplings $g_{MB}$, as is evident from Tab.\,\ref{tab:gmb}.
The set $(D,F)$ relates to the more common pair $(\xi,g_A)$ through $\xi=F/D$
and $g_A=D+F$.
The two main attractive features, from our point of view, are that this formula
describes how the complete set of baryon octet states varies as a function of
the meson masses, and that there is a chance that it remains adequate up to
higher pion masses than is usual in the HBXPT approach (say up to
$\Mpi\sim400\!-\!500\MeV$ rather than $\Mpi\sim200\!-\!300\MeV$).
In full glory it reads
\bea
M_N&=&M_0-2(2b_0+b_D+b_F)\Mpi^2-2(b_0+b_D-b_F)\Mss^2
\nonumber
\\
&&
+\frac{g_{\pi N}}{\Fpi^2}H(\Mpi^2)
+\frac{g_{K   N}}{\Fka^2}H(\Mka^2)
+\frac{g_{\et N}}{\Fet^2}H(\Met^2)
+4\de_\pi\Mpi^4+4\de_{\bar{s}s}\Mss^4
\label{hemmert1}
\\
M_\Lambda&=&M_0-4(b_0+b_D/3)\Mpi^2-2(b_0+4b_D/3)\Mss^2
\nonumber
\\
&&
+\frac{g_{\pi \Lambda}}{\Fpi^2}H(\Mpi^2)
+\frac{g_{K   \Lambda}}{\Fka^2}H(\Mka^2)
+\frac{g_{\et \Lambda}}{\Fet^2}H(\Met^2)
+4\de_\pi\Mpi^4+4\de_{\bar{s}s}\Mss^4
\label{hemmert2}
\\
M_\Sigma&=&M_0-4(b_0+b_D)\Mpi^2-2b_0\Mss^2
\nonumber
\\
&&
+\frac{g_{\pi \Sigma}}{\Fpi^2}H(\Mpi^2)
+\frac{g_{K   \Sigma}}{\Fka^2}H(\Mka^2)
+\frac{g_{\et \Sigma}}{\Fet^2}H(\Met^2)
+4\de_\pi\Mpi^4+4\de_{\bar{s}s}\Mss^4
\label{hemmert3}
\\
M_\Xi&=&M_0-2(2b_0+b_D-b_F)\Mpi^2-2(b_0+b_D+b_F)\Mss^2
\nonumber
\\
&&
+\frac{g_{\pi \Xi}}{\Fpi^2}H(\Mpi^2)
+\frac{g_{K   \Xi}}{\Fka^2}H(\Mka^2)
+\frac{g_{\et \Xi}}{\Fet^2}H(\Met^2)
+4\de_\pi\Mpi^4+4\de_{\bar{s}s}\Mss^4
\label{hemmert4}
\eea
with all meson-baryon couplings $g_{MB}$ parametrized by only two constants,
as indicated in Tab.\,\ref{tab:gmb}.
In total this gives a parametrization with 8 unknowns:
$M_0$, $b_0$, $b_D$, $b_F$, $\xi$, $g_A$, $\de_\pi$, $\de_{\bar{s}s}$.
Note that the coefficients in front of the $\Mpi^4$ and $\Mss^4$ contributions
are common to all octet members.
To the order we are working at only two of the three pseudo-Goldstone boson
masses are linearly independent; they are connected through the Gell-Mann-Okubo
relation $3\Met^2=4\Mka^2-\Mpi^2$.
Accordingly, it makes sense to consider $(\Mpi^2,\Mss^2)$ the basic mass
coordinates.

\section{First experiences with CBXPT extrapolations}

Before applying formula (\ref{hemmert1}-\ref{hemmert4}) to our data, let us
make a few practical comments.
First, it is important to notice that this formula builds on $SU(3)$ chiral
symmetry.
Accordingly, $M_0$ denotes the (common) mass of the baryon octet in the
3-flavor chiral limit.
Phenomenology suggests a value in the range $M_0\sim770\MeV$, although with a
large error margin \cite{Meissner:1997yj}.
Clearly, this is an opportunity for the lattice to come up with a considerably
more precise determination.
Next, the coefficients $D$ and $F$ are reasonably well known from
phenomenology.
It makes sense to fix the combination $g_A=D+F=1.2694(28)$
\cite{Nakamura:2010zzi} to its value at the physical nucleon mass, as the
difference to the value in the chiral limit is expected to be small.
For the ratio $\xi=F/D$ the situation is less clear; there are two preferred
scenarios in the literature, $\xi=2/3$ and $\xi\simeq0.5$.

\begin{figure}[p]
\centering
\includegraphics[width=0.82\textwidth]{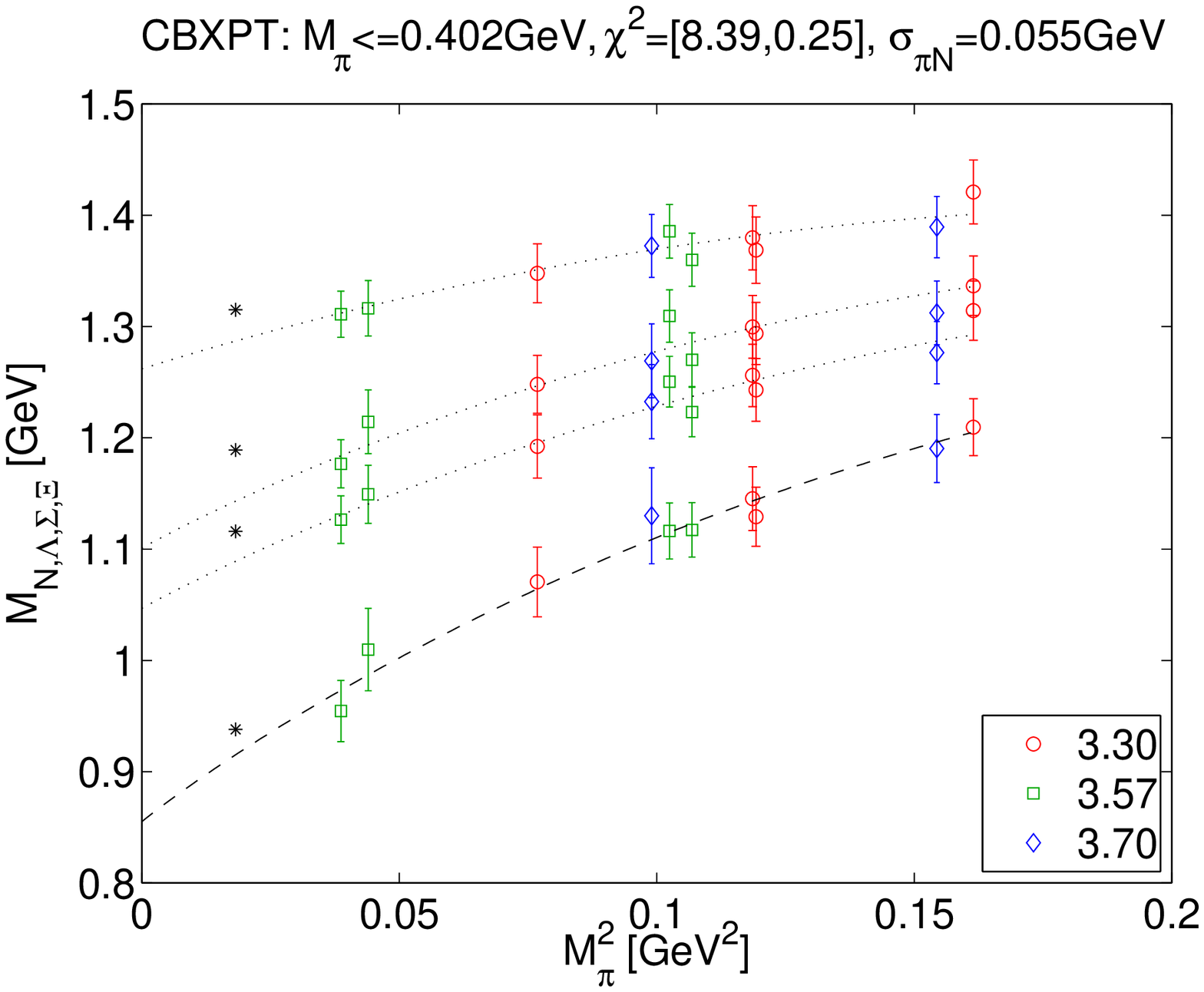}
\includegraphics[width=0.82\textwidth]{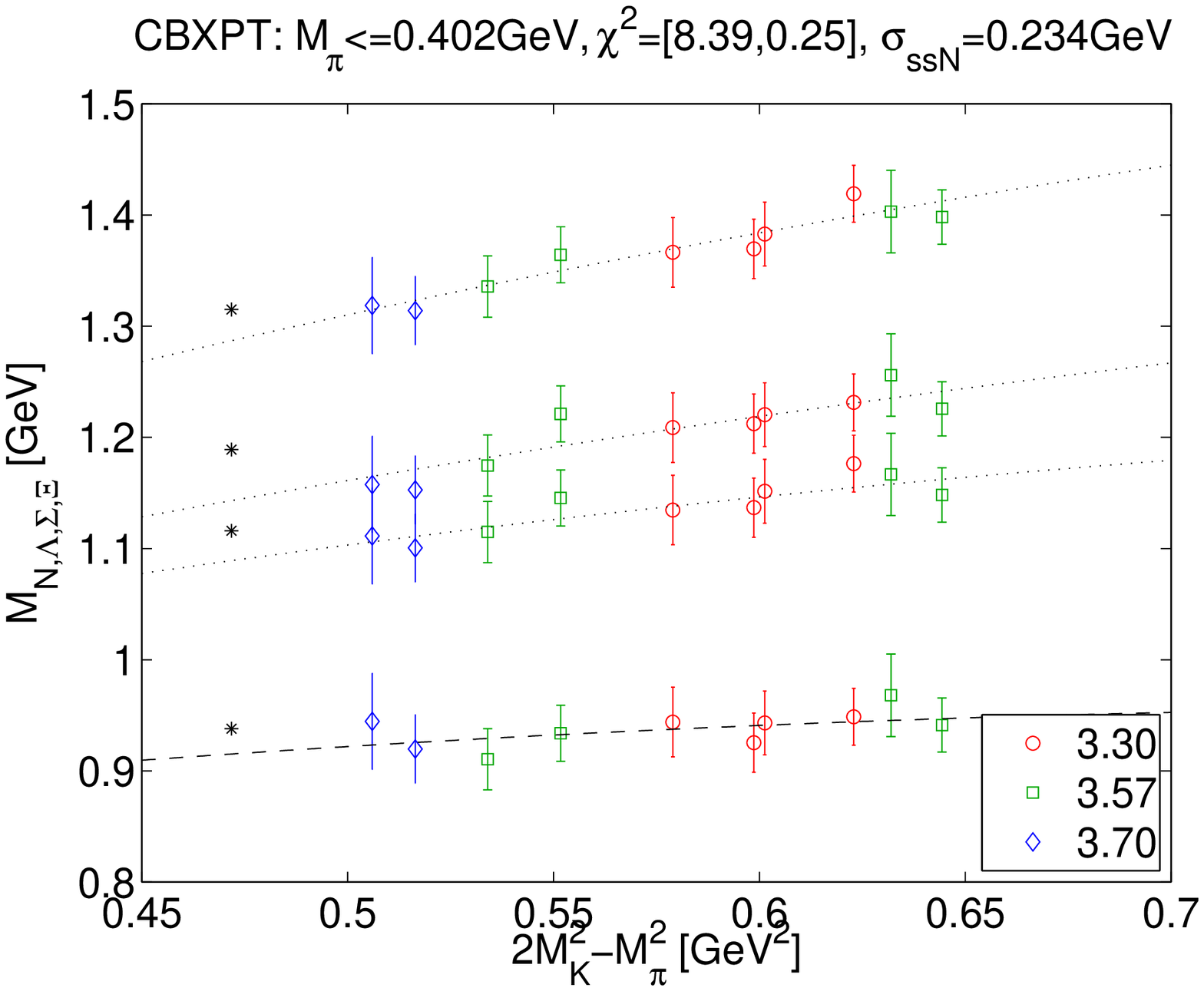}
\caption{Result of one particular ``snapshot fit'' of the ansatz (3.2-3.5) to
our ``6\,stout'' dataset with a mass cut $\Mpi<410\MeV$. For display purposes
the data have been shifted, by means of (4.1), to $[\Mss^2]_\mr{phys}$ (top)
and $[\Mpi^2]_\mr{phys}$ (bottom), and only the remaining dependence on
$\Mpi^2$ (top) or $\Mss^2$ (bottom) is shown.}
\label{fig:hemmert}
\end{figure}

A few comments are in order regarding the pseudoscalar decay constants
$\Fpi,\Fka,\Fet$.
First, let us clarify that we use the ``Bernese'' normalization where
$\Fpi^\mr{phys}=92.2\MeV$.
The more relevant point is that, whatever choice is made for $\Fpi,\Fka,\Fet$,
it is not supposed to destroy the $SU(3)$ chiral symmetry in the 3-flavor
chiral limit.
Accordingly, pinning $\Fpi,\Fka,\Fet$ down at their phenomenological values
is not an option.
We see three legitimate choices:
$(i)$ use a joint 3-flavor chiral limit value $F_0$,
$(ii)$ use an $SU(3)$ chiral formula (e.g.\ the one at NLO with $F_0,L_4,L_5$)
$(iii)$ use the measured data (perhaps with an $SU(3)$ compatible
interpolation, in any case this requires $Z_A$).

In the following we present the result of one particular fit based on choice
$(i)$.
This fit addresses the full baryon octet and yields, in principle, eight sigma
terms ($\si_{\pi N},\si_{\bar{s}sN}$ and ditto for $\Lambda,\Sigma,\Xi$).
Due to its preliminary nature, and because the comparison with the analytic
approach is not ready for other channels, we will just quote
$\si_{\pi N},\si_{\bar{s}sN}$.
Still, one should keep in mind that these numbers are preliminary, and the
assessment of systematic uncertainties is not yet finalized.

Having made these cautionary remarks, we are in a position to present in
Fig.\,\ref{fig:hemmert} the result of a fit of the ansatz
(\ref{hemmert1}-\ref{hemmert4}) to our data.
The parameters $M_0$, $b_0$, $b_D$, $b_F$, $\de_\pi$, $\de_{\bar{s}s}$, $F_0$
have been adjusted by the fit (to reasonable values), while
$(\xi,g_A)=(2/3,1.2694)$ was held fixed.
All together, this is a fit with $7$ free parameters to $40$ datapoints with an
uncorrelated $\ch^2=8.39$.
The resulting uncorrelated $\ch^2/\mr{d.o.f.}\simeq0.25$ seems plausible, since
the four octet masses in each ensemble being highly correlated will lead to an
underestimate of the true $\ch^2/\mr{d.o.f.}$ by about a factor four.

Note that this is the result of one ``snapshot fit'', i.e.\ with a
specific choice of the fitting window $[t_\mr{min},t_\mr{max}]$ for each state,
with one pion mass cut (here $\Mpi\!<\!410\MeV$), and so on.
This particular fit yields $\si_{\pi N}=55(10)_\mr{stat}\MeV$, where the quoted
error is only statistical, and $y_N\sim0.16$.
To get a trustworthy estimate of the systematic uncertainty, one should
consider reasonable variations over the fitting range, the pion mass cut, the
scaling behavior of cut-off terms, and the functional ansatz for the dependence
on $(\Mpi^2,\Mss^2)$, as will be briefly discussed in Sec.\,6 below.

A technical point worth mentioning is the shift recipe applied to show the
result of the fit.
As is clear from the discussion, the fit spans, for each baryon octet member, a
two-dimensional surface above the $(\Mpi^2,\Mss^2)$ coordinates depicted in
Fig.\,\ref{fig:landscape}.
However, because it is difficult to graphically display how $40$ datapoints
behave relative to $4$ surfaces, we resort to two one-dimensional plots.
We ``shift'' the data, along the surface established in the fit, to the
physical value of $\Mss^2$ and plot the remaining dependence on $\Mpi^2$.
In other words what is plotted in the first panel of Fig.\,\ref{fig:hemmert} is
\cite{Durr:2010hr}
\beq
\mr{data}(\Mpi^2,2\Mka^2\!-\!\Mpi^2)-
\mr{fit}(\Mpi^2,2\Mka^2\!-\!\Mpi^2)+
\mr{fit}(\Mpi^2,[2\Mka^2\!-\!\Mpi^2]_\mr{phys})
\label{shiftrecipe}
\eeq
and a similar shift is applied, in the second panel, to bring all datapoints
to a common value of $\Mpi^2$ and depict the remaining dependence on $\Mss^2$.
It goes without saying that the recipe (\ref{shiftrecipe}) does not affect the
fit itself, it just helps to display the result.

\section{First experiences with analytic extrapolations}

A complete analysis of a phenomenological variable at the physical mass point
must include a variation over the fitting ansatz that is used to interpolate
or extrapolate the data.
To this aim we include polynomial and rational ansaetze into our analysis, in
the same way as we did in  \cite{Durr:2010hr}.

\begin{figure}[tb]
\centering
\includegraphics[width=0.82\textwidth]{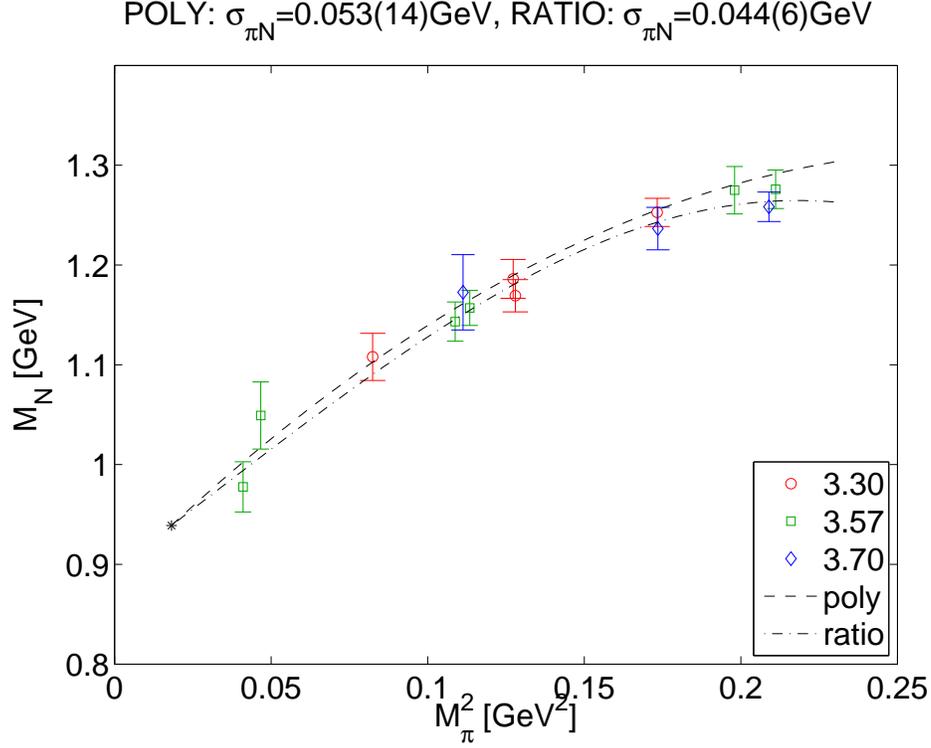}
\caption{Result of one particular ``snapshot fit'' with the polynomial and
rational ansatz (5.3,\,5.4), where the scale is set through the $N$. The
same shift procedure (4.1) has been applied to the data as in Fig.\,2.}
\label{fig:albertoplot}
\end{figure}

\begin{figure}[p]
\centering
\includegraphics[width=0.82\textwidth]{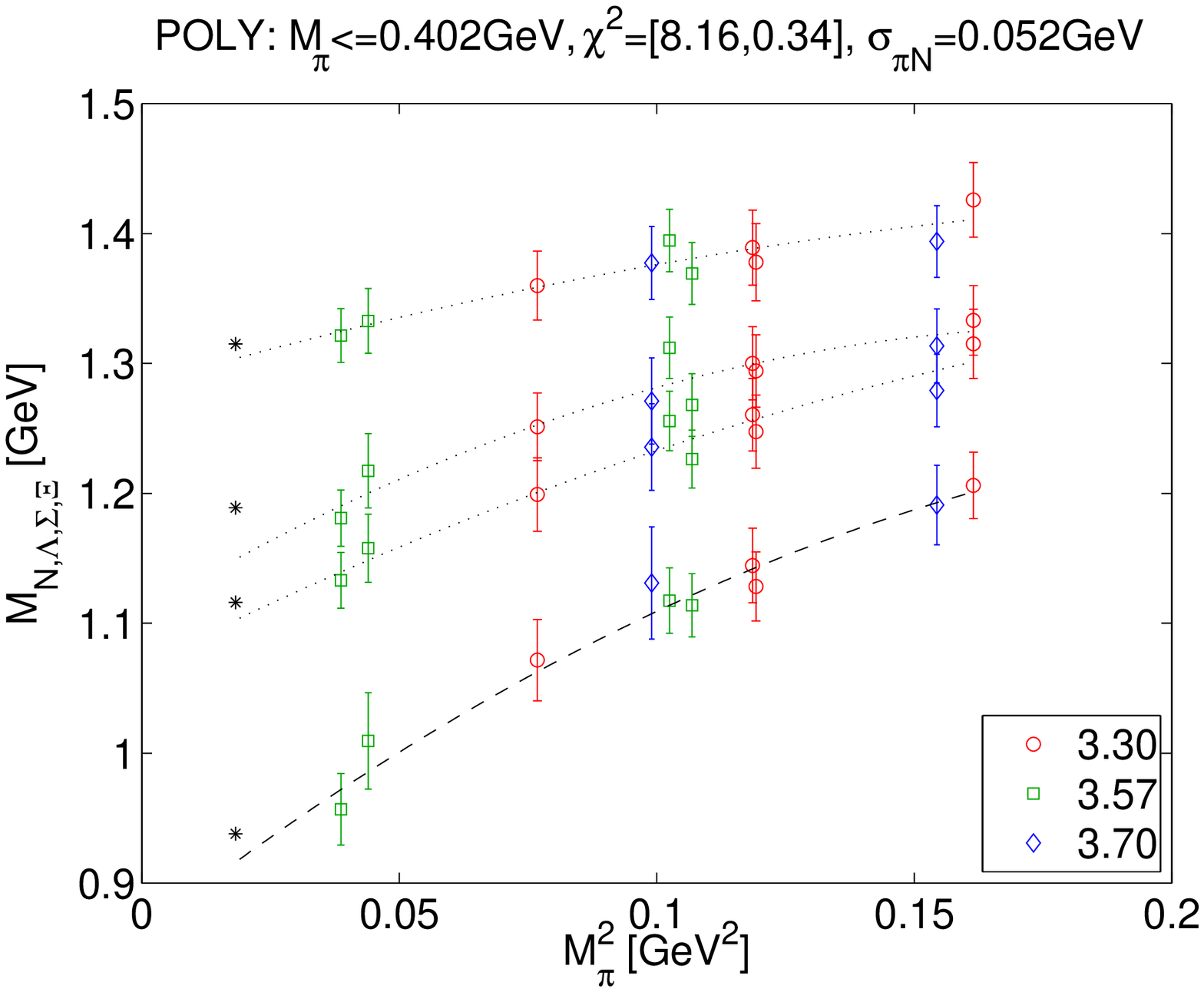}
\includegraphics[width=0.82\textwidth]{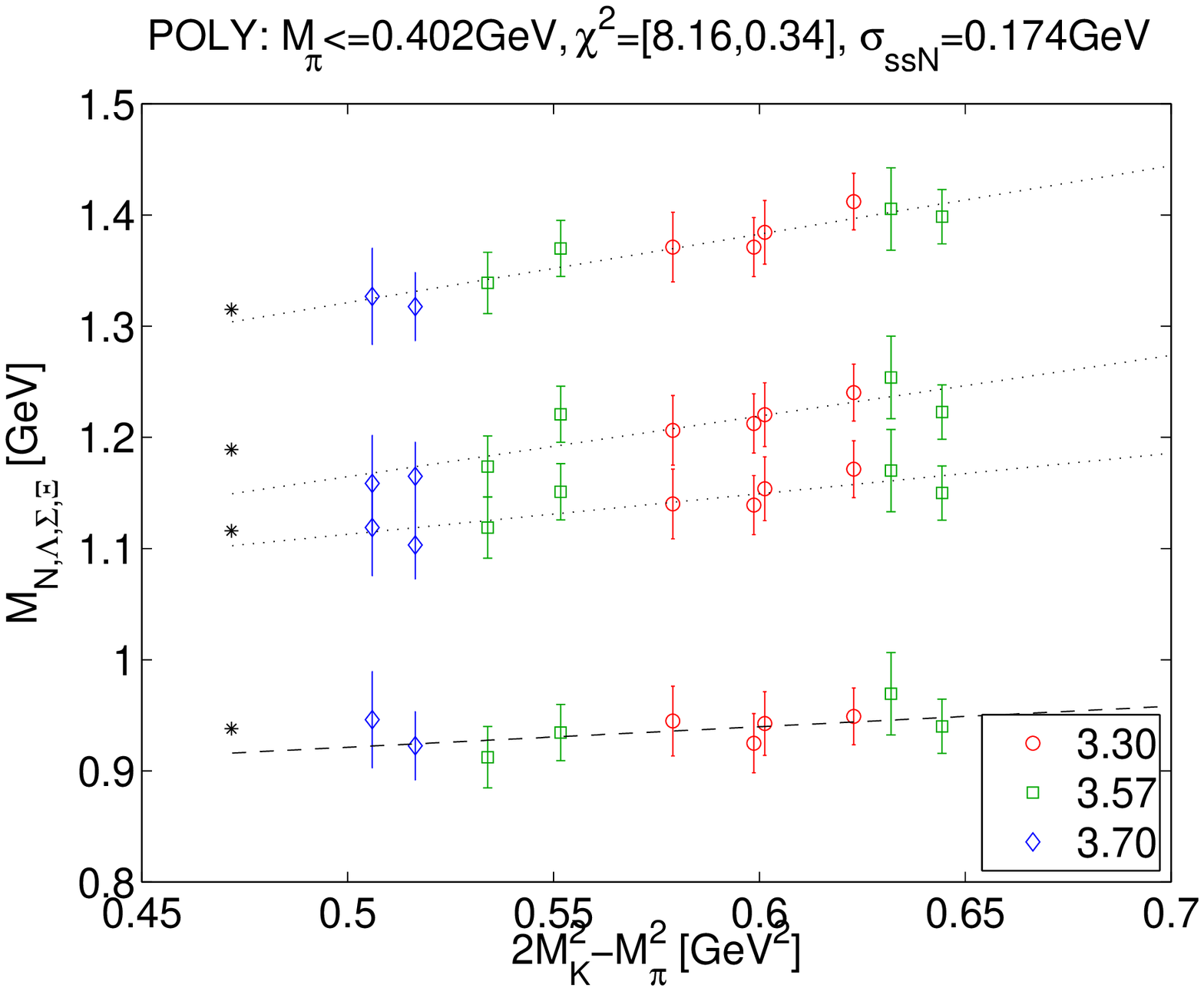}
\caption{Result of one particular ``snapshot fit'' with the global polynomial
ansatz, where the scale is set through the $\Omega$, with $\Mpi<410\MeV$. The
same shift procedure (4.1) has been applied to the data as in Fig.\,2.}
\label{fig:polynomial}
\end{figure}

This analytic approach builds on the fact that an expansion of QCD Green's
functions about the physical point $(m_{ud}^\mr{phys},m_s^\mr{phys})$ is
completely regular.
Therefore it makes sense to define dimensionless expansion parameters
$\Delta_\pi\sim(m_{ud}\!-\!m_{ud}^\mr{phys})/\Lambda$,
$\Delta_{\bar{s}s}\sim(m_s\!-\!m_s^\mr{phys})/\Lambda$, and to express the
measured octet mass as a function of these parameters.
At this point details matter.
We prefer
\beq
\Delta_\pi=
\Big(\frac{a\Mpi}{a\!\cdot\!M_X^\mr{phys}}\Big)^2-
\Big(\frac{\Mpi^\mr{phys}}{\!M_X^\mr{phys}}\Big)^2
\quad,\qquad
\Delta_{\bar{s}s}=
\Big(\frac{a\Mss}{a\!\cdot\!M_X^\mr{phys}}\Big)^2-
\Big(\frac{\Mss^\mr{phys}}{\!M_X^\mr{phys}}\Big)^2
\label{delta_one}
\eeq
with $X$ being the scale setting state over the alternative choice
\beq
\Delta_\pi'=
\Big(\frac{a\Mpi}{aM_X}\Big)^2-
\Big(\frac{\Mpi^\mr{phys}}{\!M_X^\mr{phys}}\Big)^2
\quad,\qquad
\Delta_{\bar{s}s}'=
\Big(\frac{a\Mss}{aM_X}\Big)^2-
\Big(\frac{\Mss^\mr{phys}}{\!M_X^\mr{phys}}\Big)^2
\label{delta_two}
\eeq
since in (\ref{delta_one}) all dependence on the simulated $am_{ud}$, $am_s$ is
in the very first numerator, while in (\ref{delta_two}) the numerator and the
denominator of the first terms depend on the quark masses.

With such small mass parameters in hand, and bearing in mind that the range
of simulated $\Mpi^2$ is much larger than the range of simulated $\Mss^2$
(cf.\ Fig.\,\ref{fig:landscape}), we consider the polynomial ansatz
\beq
(aM_N)=a\!\cdot\!M_N^\mr{phys}
\Big[1+c_1\Delta_\pi+c_2\Delta_\pi^2+c_3\Delta_{\bar{s}s}\Big]
\label{ansatz_poly}
\eeq
where $c_1,c_2,c_3$ and the isolated $a$ (per coupling $\beta$, cf.\ the
discussion in Sec.\,2) are the fit parameters.
Obviously, with this choice one sets the scale through the nucleon mass, which
means that the quantity that is effectively calculated is $\si_{\pi N}/M_N$ at
the physical mass point.
Likewise
\beq
(aM_N)=a\!\cdot\!M_N^\mr{phys}
\Big[1-d_1\Delta_\pi-d_2\Delta_\pi^2-d_3\Delta_{\bar{s}s}\Big]^{-1}
\label{ansatz_ratio}
\eeq
is a rational ansatz with similar characteristics.
With any such fit in hand, one proceeds along the lines of
(\ref{factorization}) and evaluates the change of the fit function under a
change of $(a\Mpi)^2$, $(a\Mss)^2$.
With a simple ansatz like (\ref{ansatz_poly}) or (\ref{ansatz_ratio}) this can
even be done analytically, giving
\bea
\si_{\pi N}&=&m_{ud}\frac{\pad M_N}{\pad m_{ud}}\Big|_\mr{phys}=
\Mpi^2\frac{\pad M_N}{\pad\Mpi^2}=\Big[\frac{\Mpi^2}{M_N}\Big]_\mr{phys}c_1
\\
\si_{\bar{s}sN}&=&2m_s\frac{\pad M_N}{\pad m_s}\Big|_\mr{phys}=
2\Mss^2\frac{\pad M_N}{\pad\Mss^2}=2\Big[\frac{\Mss^2}{M_N}\Big]_\mr{phys}c_3
\\
y&=&\frac{m_s}{m_{ud}}\frac{\si_{\bar{s}sN}}{\si_{\pi N}}\Big|_\mr{phys}=
2c_3/c_1
\;.
\eea

In the same spirit, one may consider a global polynomial or rational ansatz
which fits the entire baryon octet at once.
In this case setting the scale through an octet member would introduce some
asymmetry, and it seems advisable to include one more fit to use $X\!=\!\Omega$
for this purpose.

With the same cautionary remarks applicable as in Sec.\,4, we show some
``snapshot fits'' to the ansaetze (\ref{ansatz_poly}) and (\ref{ansatz_ratio})
in Fig.\,\ref{fig:albertoplot}.
They yield $\si_{\pi N}=53(14)_\mr{stat}\MeV$ and
$\si_{\pi N}=44(6)_\mr{stat}\MeV$, respectively, whith the quoted errors being
statistical only.
The result of a joint ``snapshot fit'' to the full octet is presented in
Fig.\,\ref{fig:polynomial}.
Here, the scale is set through the $\Omega$, and the same pion mass cut
$\Mpi\!<\!410\MeV$ is used as in Fig.\,\ref{fig:hemmert}.
Both $\si_{\pi N}=52(10)_\mr{stat}\MeV$ and $y_N\sim0.13$ are well compatible
with what was found in the CBXPT approach.
In fact, with either fit the strangeness content is well consistent with zero.
Going back to Fig.\,\ref{fig:landscape} and the bottom panel of
Fig.\,\ref{fig:hemmert} or Fig.\,\ref{fig:polynomial} it appears that this is
linked to our strange masses being slightly above the physical target value.
Likely, with simulated values of $2\Mka^2\!-\!\Mpi^2$ covering the entire
range between (say) $0.3\GeV^2$ and $0.65\GeV^2$ [such that the physical
value of $0.47\GeV^2$ would be the center of this range] one would have a
better chance of determining the slope.
Note that for the other octet members, the situaton looks better, since their
slope in the $\bar{s}s$ direction seems to be larger.

\section{Preliminary results and outlook}

Given the discussion in the previous two sections, it is clear how we should
proceed to come up with a valid determination of the nucleon sigma terms (or
equivalently of $\si_{\pi N}$, $y_N$) and analogous sigma terms for the baryon
octet states.

We have implemented several functional ansaetze to describe the dependence
of the baryon state on $(\Mpi^2,\Mss^2)$.
Both the CBXPT ansatz tested in Sec.\,4 and the family of polynomial and
rational ansaetze tested in Sec.\,5 yield reasonable results for the sigma
terms (and in the former case also for the low-energy parameters).
With such an interpolation in hand, one may compute the derivatives with
respect to $\Mpi^2$, $\Mss^2$, at the physical mass point.
In some cases this is a simple function of the fitted parameters.
In other cases it proves more convenient to evaluate the derivatives with
respect to the measured $\Mpi$, $\Mka$ and to convert via
\bea
\si_{\pi N}(m_{ud})\simeq
\Mpi^2\frac{dM_N}{d\Mpi^2}\Big|_{\Mss\,\mr{fixed}}&=&
\Mpi^2\frac{\pad M_N(\Mpi,\Mka)}{\pad\Mpi^2}+
\Mpi^2\frac{\pad M_N(\Mpi,\Mka)}{\pad\Mka^2}
\frac{\pad\Mka^2}{\pad\Mpi^2}\Big|_{\Mss\,\mr{fixed}}
\nonumber
\\
&=&
\frac{\Mpi}{2}\frac{\pad M_N(\Mpi,\Mka)}{\pad\Mpi}+
\frac{\Mpi^2}{4\Mka}\frac{\pad M_N(\Mpi,\Mka)}{\pad\Mka}
\\
\si_{\bar{s}sN}(m_s)\simeq
2\Mss^2\frac{dM_N}{d\Mss^2}\Big|_{\Mpi\,\mr{fixed}}&=&
2\Mss^2\frac{\pad M_N(\Mpi,\Mka)}{\pad\Mka^2}
\frac{\pad\Mka^2}{\pad\Mss^2}\Big|_{\Mpi\,\mr{fixed}}
\nonumber
\\
&=&
\frac{\Mss^2}{2\Mka}
\frac{\pad M_N(\Mpi,\Mka)}{\pad\Mka}
\eea
where again everything is evaluated at the physical mass point.

What remains to be done is the systematic variation of all ansaetze over the
original fitting window, over the pion mass cut, and over the scaling pattern
of potential discretization effects.
Of course, all of this should be done on $O(1000)$ bootstrap samples to assess
the statistical uncertainty, but this part is standard.
This kind of machinery was used in Ref.\,\cite{Durr:2010hr} to give a reliable
assessment of both the statistical and the systematic uncertainty of the
observable of interest ($f_K/f_\pi$).

For the pion-nucleon sigma term in MeV units our fits usually yield values in
the lower fifties with typically about ten MeV statistical error.
With hindsight we thus anticipate that the final result will be in the range
of $\si_{\pi N}\simeq50(10)(10)\MeV$, where in each slot only one digit is
meant to be significant.
Regarding $\si_{\bar{s}sN}$ or $y_N$ the situation is less convincing.
With the ``6\,stout'' dataset depicted in Fig.\,\ref{fig:landscape} we obtain,
with each ansatz, large statistical errors and non-negligible spreads among the
ansaetze.
Currently, a value like $y_N\simeq0.1(2)(1)$ seems appropriate, which would
not even tell whether there is a non-zero strangeness content at all.
Our current understanding suggests that, in order to obtain a substantially
more precise value, one would have to add simulation points with significantly
smaller strange quark mass than we have right now.

\acknowledgments
\noindent
This work is part of an ongoing callaboration between the universities of
Wuppertal and Regensburg in the framework of SFB TR-55.
Computations were performed using HPC resources from FZ J\"ulich and
GENCI-[CCRT/IDRIS] (grant 52275), as well as clusters at Wuppertal and CPT.
This work is supported in part by EU grants FP7/2007-2013/ERC n$^o$ 208740,
MRTN-CT-2006-035482 (FLAVIAnet), OTKA grant AT049652, DFG grant FO502/2,
SFB-TR 55, U.S.\ Department of Energy Grant No.\ DE-FG02-05ER25681, CNRS grants
GDR n$^0$ 2921 and PICS n$^0$ 4707.



\end{document}